\documentclass[journal,compsoc]{IEEEtran}

\usepackage{graphicx}
\usepackage{cite}
\usepackage[hidelinks]{hyperref}
\usepackage{todonotes}
\usepackage{rotating}

\title{Performance analysis of a Tor-like onion routing implementation}
\author{\IEEEauthorblockN{Quinten Stokkink} \quad \IEEEauthorblockN{Harmjan Treep} \quad \IEEEauthorblockN{Johan Pouwelse}}
\date{}

\begin{document}

\maketitle

\begin{abstract}
	The current onion routing implementation of Tribler works as expected but throttles the overall throughput of the Tribler system.
	This article discusses a measuring procedure to reproducibly profile the tunnel implementation so further optimizations of the tunnel community can be made.
	Our work has been integrated into the Tribler eco-system.
\end{abstract}

\section*{Introduction}
	Any sense of privacy on The Internet is an illusion.
	Even Tor, which could be considered the most privacy preserving networking framework in existence, has its flaws \cite{syverson2013practical}.
	Furthermore, privacy-enhancing technology is difficult to use for normal users and it slows down The Internet.
	For instance, should one first setup a VPN and then Tor or Tor and then VPN?
	Getting it wrong might actually negatively impact a user's privacy\footnote{\url{https://trac.torproject.org/projects/tor/wiki/doc/TorPlusVPN}}.

	With yearly sales of smartphones and smart watches approaching one billion, threats to user privacy are becoming a global phenomenon. 
	People can be traced to a location within (in the worst case) 20 meters in real-time\footnote{\url{http://buddy-locator.com/}}\footnote{\url{http://www.mobile-scan.com/}} and recognized\cite{de2013unique}.
	With these billions of people facing threats ranging from targeted advertisement to burglary or even harassment, the need for scalable and light-weight privacy-enhancing technology becomes apparent.
	However, no \emph{optimized} implementation of a scalable architecture exists in this emerging research field. We provide a key step forward by identyfing bottlenecks in Tribler.

\bigskip

	The Tor project\footnote{\url{https://www.torproject.org/}} aims to offer anonymity by forwarding traffic through a series of relays. 
	Multiple layers of encryption are utilized such that no single relay can reconstruct the entire circuit.
	This is also called onion routing.
	These relays are provided by volunteers which means there is often not enough bandwidth available causing the Tor network to be slow; almost no-one uses it for everyday browsing.

	The solution is to make everyone in the network a relay for others\cite{rplak}.
	Several implementations of such an approach are available, for instance, Tribler\footnote{\url{https://tribler.org/}} and Hola\footnote{\url{https://hola.org/}}.
	Their architecture allows them to scale to both a large number of users and high-bandwidth applications such as HD video streaming.

	The key contribution of this article is a performance analysis of the first implementation of Tor-derived onion routing implementation with user-donations, NAT/firewall puncturing and fully decentralized peer discovery: Tribler.
	To make our test realistic we procured various anonymous private servers in exotic locations such as Belize and Noord-Holland.

\section*{Problem description}
	\begin{figure}
		\centering
		\includegraphics[width=\columnwidth, trim=0cm 0cm 0cm 1.5cm, clip=true]{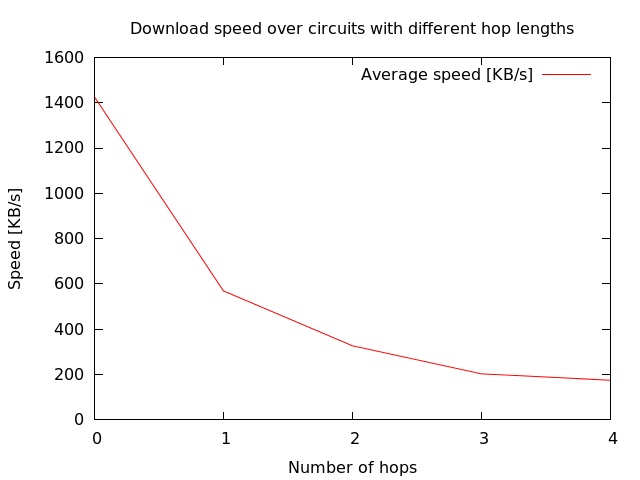}
		\caption{Download speed per amount of hops}
		\label{fig:down-per-hop}
	\end{figure}

	The current onion routing implementation can find a 2 hop path between the piece of information needed and the current computer running Tribler.
	It also allows building circuits of variable length, but this is currently not used as it severely limits the throughput of downloads.
	This was demonstrated by an experiment in 2014 among Tribler users.
	In this experiment dozens of users downloaded a 50 MB file using the tunnel code from a 1 Gbps server\footnote{\href{http://forum.tribler.org/viewtopic.php?f=2&t=2121&start=80\#p8817}{http://forum.tribler.org/viewtopic.php?f=2\&t=2121}}.
	The results of this experiment can be observed in \autoref{fig:down-per-hop}.
	However, it was unknown what the limiting factors were.
	Possible culprits included the openssl library calls, the single core processing, the IO capacity or the network bandwidth.

\section*{Experiment setup}
	The contents of this section are as follows.
	First it will explain the high-level construction of the node network in the experiment.
	Secondly the frameworks that were used are introduced.
	Lastly the limitations and practical details of our setup are discussed.

\bigskip

\begin{figure}
		\centering
		\includegraphics[width=\columnwidth]{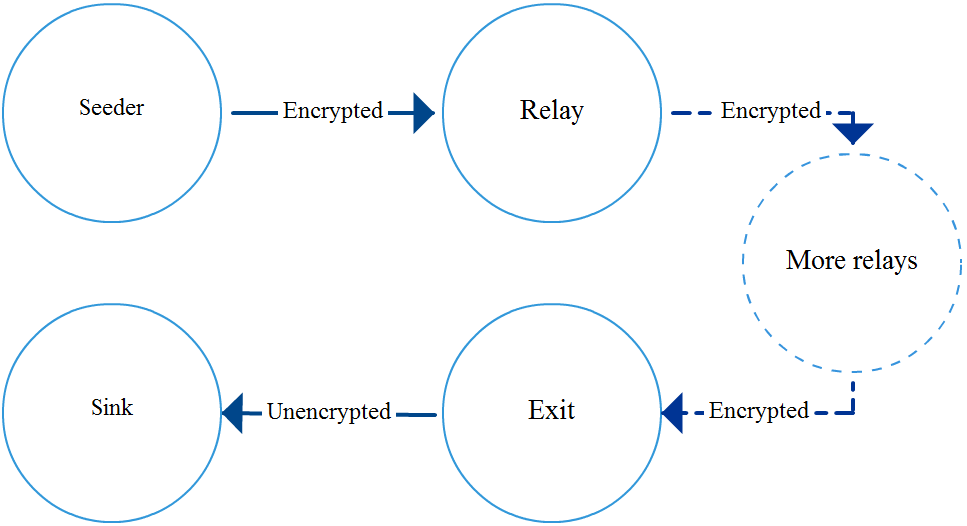}
		\caption{Experiment setup}
		\label{fig:experiment-setup}
	\end{figure}

	In this experiment a small community of nodes is created.
	An example of one such circuit in this network is shown in \autoref{fig:experiment-setup} and consists of a:

	\begin{itemize}
		\item Seeding node, sending data
		\item Exit node, relaying the unencrypted data to the sink
		\item Sink node, receiving all of the data from the Seeding node
		\item Relay node, relaying encrypted packets within a circuit
	\end{itemize}

	An experiment run consists of the seeding node constructing 4 different circuits using 0 to 3 hops (relay nodes + 1 exit node), to send to the sink node.
	Each of these circuits are built and destroyed independently of each other.
	After building the circuits, some random data is sent between the first and last node in the circuit.
	The circuit with 0 hops just encrypts and decrypts information on the same node.
	This allows us to run the experiment without network overhead.
	At the end of the run the results are evaluated per node type.

\bigskip

	Dispersy \cite{zeilemaker2014large} allows for decentralized communities of nodes to communicate using custom protocols.
	The tunnel community in Tribler is one of these Dispery communities.
	Through the implementation of the tunnel community, nodes can announce, discover and share candidate exit nodes and relays.
	In the experiment of this paper a single instance of the Tribler tunnel community is created without the support of other Tribler functionalities.

\bigskip

		Gumby is the experiment runner for the Tribler project.
		It allows for creating repeatable tests that can spawn instances on different computer setups, for example running all Tribler instances on different DAS4 computers or running it all locally.
		To provide this functionality it uses high level scenarios which interact with python boilerplate code to interact with core Tribler functionality.
		This paper has expanded upon Gumby by creating a scenario for testing the tunnel implementation with various numbers of hops and circuits and providing the python boilerplate code for running this scenario using either random packet transmission or LibTorrent controlled packet transmission over the tunnel's circuits.

\bigskip

		Yappi\footnote{\url{https://code.google.com/p/yappi/}} is the profiler used to obtain the profiling information from the application while the application is running.
		Yappi is designed to support multi-threaded applications and be started and stopped without affecting the application, which makes it the suited debugger for the Tribler components.
		Depending on the configuration, it returns the amount times a function was called and either the wall time or cpu time per function.

		To understand the timing of the tunnel component we utilized Yappi's CPU time reporting and filtered the results to only include function calls related to the tunnel implementation. 
		These results per function are then sorted and plotted relatively in a pie chart and plotted absolutely in a bar graph using R scripts.

\bigskip

	Due to budget and time constraints the experiment has only been run on a single machine, running 8 instances of Tribler.
	This approach places extra strain on the system resources, but this should be of no consequence when interpreting the relative results of function performance.
	This also means that packets do not travel over the wild internet in the experiment (instead they bounce back from the on-site router).
	However, delays imposed due to packets having a longer transmission time should not matter for the performance of the implementation.
	In fact, if these transmission delays impact the time spent in functions, one could argue that this does not measure the core performance of the function and thus gives a skewed result of its performance.

\section*{Experimental results}
\begin{figure}
		\centering
		\includegraphics[width=\columnwidth]{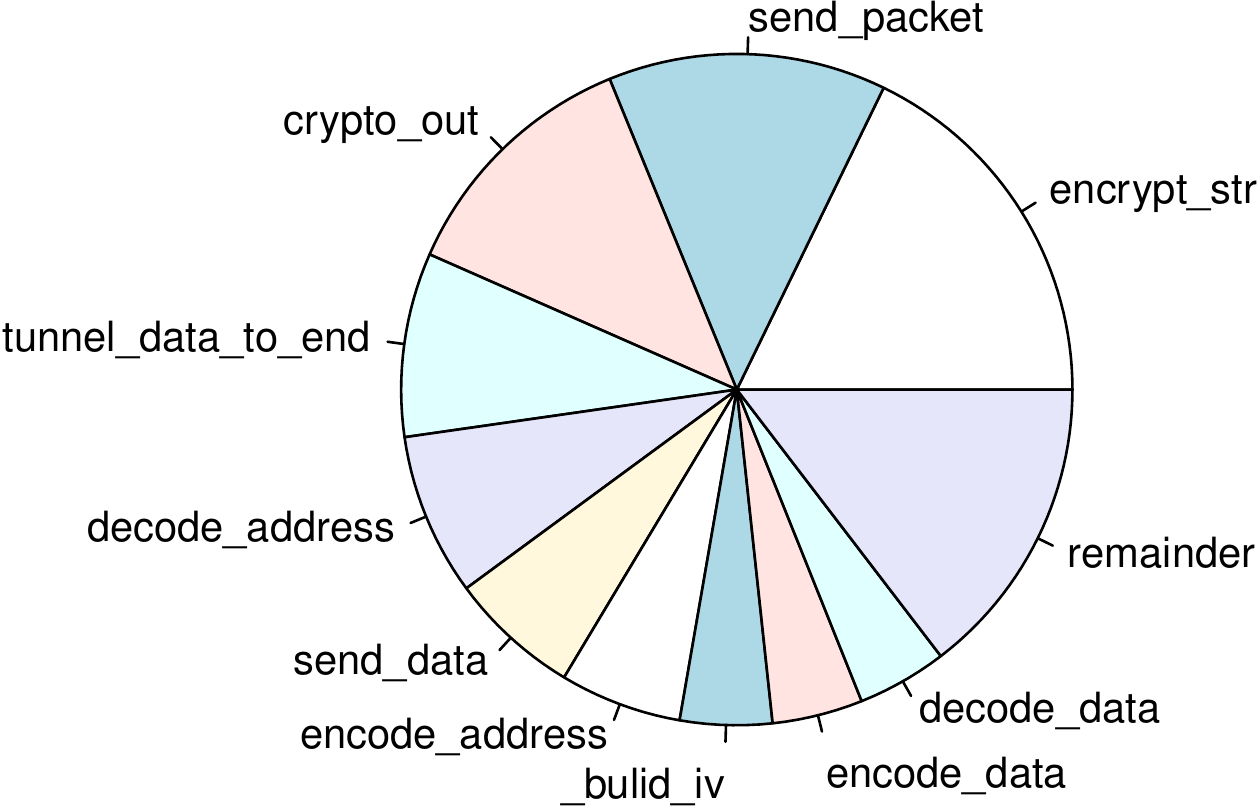}
		\caption{Relative CPU time per function of the seeding node for 0 hops and 4 circuits}
		\label{fig:result-seed-0}
	\end{figure}

\begin{figure}
		\hspace*{.39in}
		\centering
		\includegraphics[width=0.9\columnwidth]{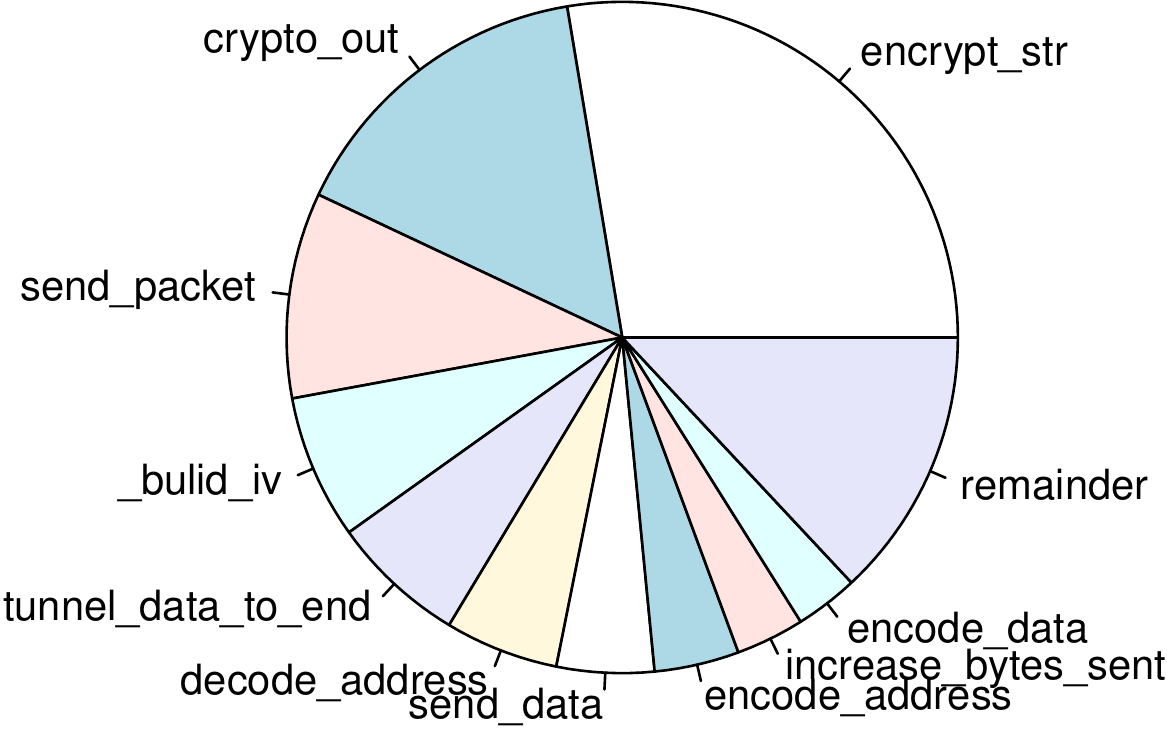}
		\caption{Relative CPU time per function of the seeding node for 3 hops and 4 circuits}
		\label{fig:result-seed-3}
	\end{figure}

	The initial observervation to be made when interpreting the results is the fact that the seeding node is influenced the most by the increasing number of hops.
	As the number of hops increased per run, the cryptographic component of the seeding node (\textit{encrypt\_str} and \textit{crypto\_out}) started to take a greater toll on its performance (see \autoref{fig:result-seed-0} and \autoref{fig:result-seed-3}).
	This is in contrast to the relay and exit nodes which appeared to not differ at all (performance wise) as the number of hops increased.
	This result is in line with the onion routing model, as the seeding node is in charge of encrypting a packet multiple times, for each hop in a circuit.
	The exit and relay nodes only need to decrypt this packet once and forward it to another node.


	To better explain and distinguish the different results we shall categorize the functions into two seperate functionally similar groups.
	This is mainly done to provide abstraction over the circuit implementation, where functions might perform the same functionality but in a different setting (like \textit{send\_packet} for the seeding node and \textit{relay\_packet} for the exit/relay nodes).
	The first group we distinguish is the \textbf{crypto} set.
	This group contains the functions that pertain to encryption or decryption of messages.
	The functions contained in this first category are the \textit{encrypt\_str}, \textit{decrypt\_str}, \textit{encode\_address}, \textit{decode\_address} and \textit{crypto\_out} functions.
	The second group is the \textbf{networking} set.
	This group contains functions that handle creating and sending the actual UDP packages over the internet.
	The functions of this category are the \textit{send\_packet} and \textit{relay\_packet} functions.
	All of the other functions we assign to the \textbf{other} set.

	\begin{table}
		\centering
		\begin{tabular}{l||c|c|c}
			                         & 0 Hops & 3 Hops & Exit \\ \hline
			\textit{encrypt\_str}    & 0.18 & 0.28 & 0.00 \\
			\textit{decrypt\_str}    & 0.00 & 0.00 & 0.15 \\
			\textit{encode\_address} & 0.06 & 0.04 & 0.00 \\
			\textit{decode\_address} & 0.08 & 0.05 & 0.05 \\
			\textit{crypto\_out}     & 0.12 & 0.15 & 0.00 \\ \hline
			                         & 0.44 & 0.53 & 0.20
		\end{tabular}
		\caption{The relative processing times of the crypto set of functions}
		\label{tbl:cryptoset}
	\end{table}
	
	\begin{table}
		\centering
		\begin{tabular}{l||c|c|c}
			                       & 0 Hops & 3 Hops & Exit \\ \hline
			\textit{send\_packet}  & 0.13 & 0.10 & 0.08 \\
			\textit{relay\_packet} & 0.00 & 0.00 & 0.11 \\ \hline
			                       & 0.13 & 0.10 & 0.19
		\end{tabular}
		\caption{The relative processing times of the network set of functions}
		\label{tbl:netset}
	\end{table}
	
	What can be observed when analyzing this data is that the \textbf{crypto} set's functions take up the most time.
	The secondmost impactful functions are those of the \textbf{networking} set.
	As seen in \autoref{tbl:cryptoset} and \autoref{tbl:netset} these two sets of functions take more than 50\% of the CPU time for tunnel community code in the seeding experiment.
	In the exit node they use almost 40\% of the CPU time.
	From this we can conclude that the tunnel components which would benefit most of parallellization are these two classes of functions.
	Also we note that this is not an easy task since these two classes are mostly reliant on external libraries in Tribler's implementation.
	Parallelizing functions in the \textbf{other} set might not be a good idea though as the parallelization overhead might outweigh the execution times of the functions.
	This is of course dependent on the machine Tribler is being executed on, but if one considers smartphones as the targeted platform for optimization, this would indeed be a bad idea.

	From \autoref{fig:result-seed-0} and \autoref{fig:result-seed-3} it can be observed that the \textit{encode\_address} and \textit{decode\_address} take a lot of CPU time.
	After manual inspection it has been determined that the \textit{encode\_address} function is concerned with converting IP strings to a binary format.
	Because of their repeated use, these results can be cached to save time.
	One way to do this, would be to save the converted IP strings in their binary format.
	When \textit{decode\_address} is observed, we see that this method could also be used for reverse lookups of host and port tuples.
	Better yet, for both functions, would be to only work with the binary encoded addresses in the entirety of the tunnel community and only convert them to string format when it is really needed (like user interactions).

	Lastly we can look to optimizations that can be done to speed up the onion routing encryption, which responds the worst to scaling up the amount of hops.
	This is not something that can be fixed by using normal simple parallelization however, but due to the fact that onion routing requires successive RSA encryptions on the same packet.
	The only way to speed this up, is to perform pipelined header encryption.
	However, this would require a great amount of control over Python's packet compression and encryption library.
	Even though this would aid performance, most of a normal PC's resources will already be in use thanks to the parallelism of the different circuits being used.

\bigskip

	On a more course grained scope of parallellization we have found, after inspecting the source code of Tribler, that sending or relaying occurs sequentially after the encryption and decryption of a packet.
	If this process was pipelined perfectly, the results show a speedup to 15.46\% for the 0 hop and 10.99\% for the 3 hop seeding node experiment and a speedup of 12.5\% for the exit node experiments.
	This pipelining could be achieved by usage of transmission buffers or a thread-safe double ended queue, which store encrypted packets awaiting to be sent by another thread.
	In contrast to the per-function parallellization this would be relatively trivial to implement, but a significant improvement nonetheless.

\bigskip

	One quirk in the results is the total amount of time spent in the functions of the seeding node.
	The experiments showed that as the number of hops increased, the total time spent in the functions of the seeding node decreased.
	It is speculated that this is due to some form of load balancing among the nodes.
	This would be due to the node processes being allocated to different cores on the simulating machine.
	The exact remains a mystery and a target for future work.

\bigskip
	
\begin{figure*}
	\centering
	\includegraphics[width=\textwidth]{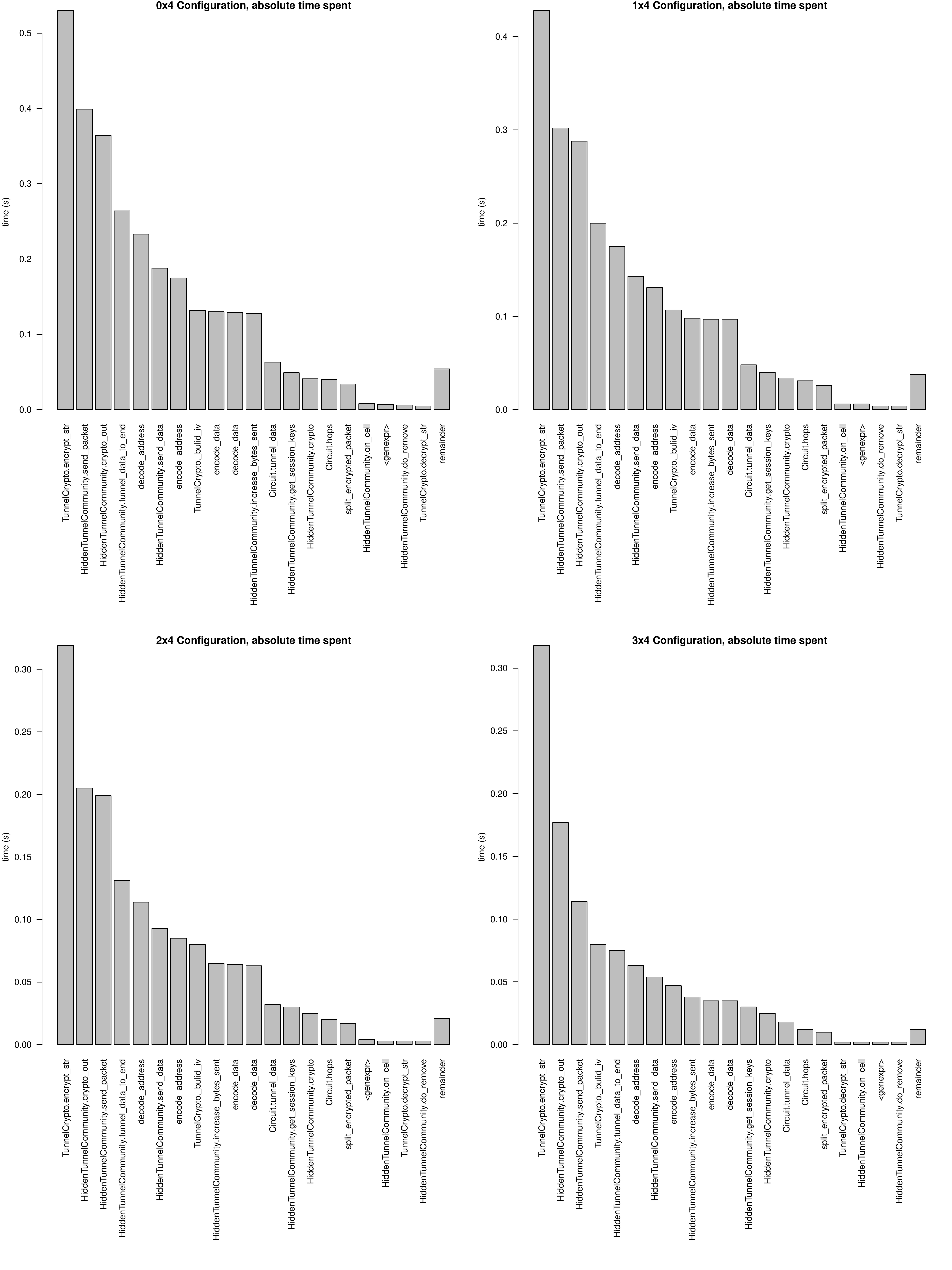}
	\caption{Absolute time spent for different (\#hops, \#circuits) experiments for the seeding node}
	\label{fig:bar-plots-seed}
\end{figure*}

\begin{figure*}
	\centering
	\includegraphics[width=\textwidth]{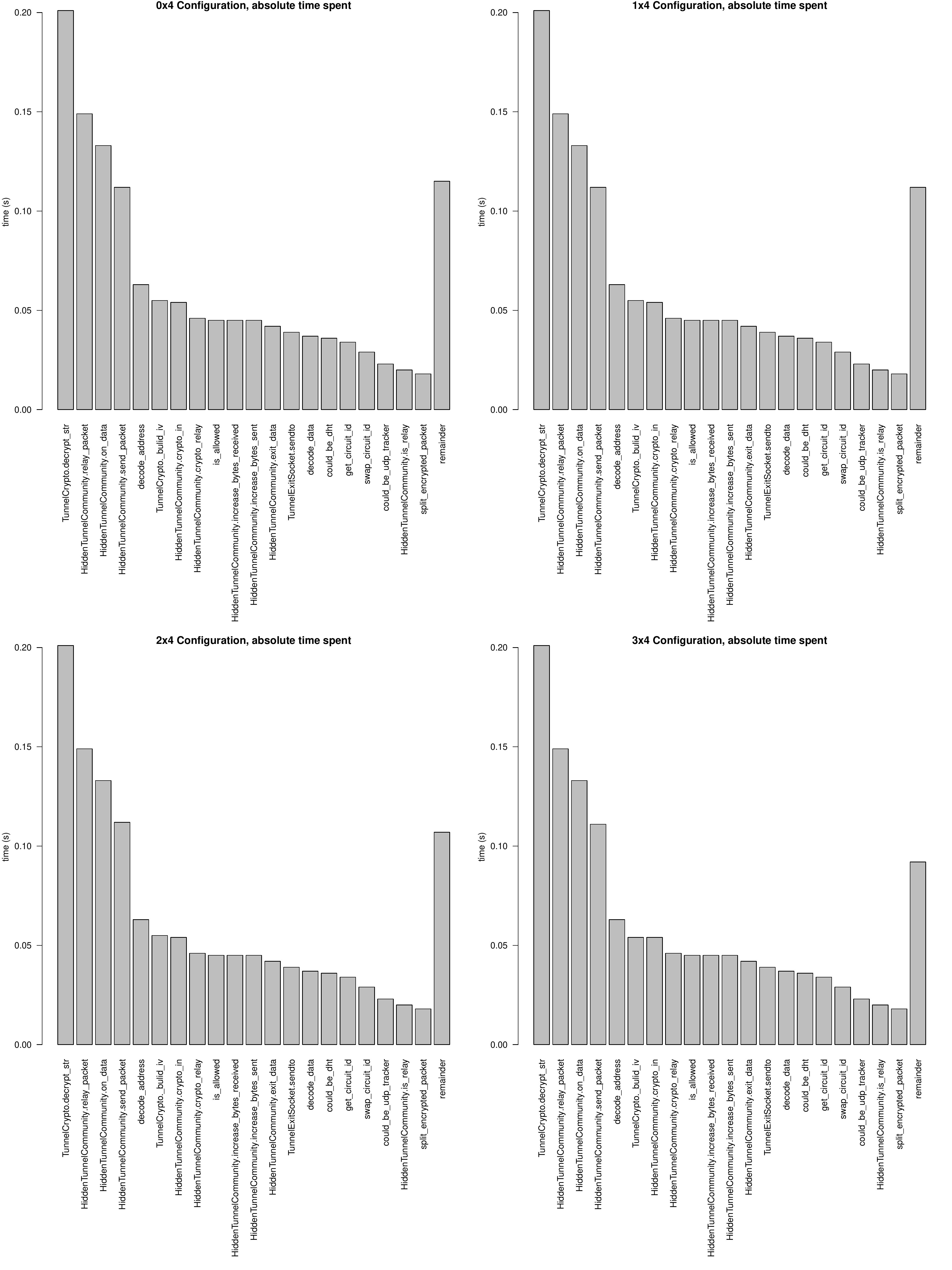}
	\caption{Absolute time spent for different (\#hops, \#circuits) experiments for the exit node}
	\label{fig:bar-plots-exit}
\end{figure*}

	The exact total runtimes for the top 20 functions in the seeding node and exit node experiments can also be found in \autoref{fig:bar-plots-seed} and \autoref{fig:bar-plots-exit} respectively.

\section*{Future work}
	The next step in this research would be to utilize LibTorrent to send packets over the circuits.
	This would enable our experiment to utilize rate limiting and measure packet loss overhead.
	At the moment our implementation does not consider packet loss, in other words it is sending UDP packets blindly.
	This means packets that are received by the sink node are being dropped instantly and common occurrences such as packet retransmissions are not measured.
	One thing to keep in mind, is that using LibTorrent should not offer any new insights into throughput loss.
	This is because LibTorrent is concerned with the contents of the data that is being sent over the circuits and not how the tunnel implementation sends it over the network.
	The measurements performed by the experiment in this paper are only on the actual tunnel implementation and not on the functions that handle the received data.
	Thus, whereas this might change the absolute time spent in different functions, the relative time spent should remain the same.

\section*{Conclusion}
	In this paper we have evaluated the performance of the implementation of a privacy preserving communication protocol in a maturing application called Tribler.
	We have successfully tapped into and analyzed function calls in this implementation, both manually and using measuring frameworks.
	Our results have uncovered two sets of functions where there is major room for improvement by way of different forms of parallelization.
	We have found that these sets consists of functions concerned with (1) sending packets and (2) encrypting packets.
	Furthemore we found that these two categories of bottleneck functions are of more or less equal size, with the encryption of packets taking only slightly longer than the sending of packets.
	Our work has been made a standard part of the Tribler ecosystem and will be used for further optimizations like multi-core support.

\bibliography{bibliography}{}
\bibliographystyle{plain}

\newpage

\end{document}